\documentclass[manuscript]{aastex}

\shorttitle{Rossby}
\shortauthors{Zaqarashvili et al.}

\begin{document}

\title{Magnetic Rossby waves in the solar tachocline and Rieger-type periodicities}

\author{Teimuraz V. Zaqarashvili\altaffilmark{1,3},
Marc Carbonell\altaffilmark{2}, Ram\'{o}n Oliver\altaffilmark{3}, and Jos\'{e} Luis
Ballester\altaffilmark{3}}

\altaffiltext{1}{Abastumani Astrophysical Observatory at Faculty of Physics and Mathematics,\\
Ilia Chavchavadze State University, Chavchavadze Ave 32, 0179
Tbilisi, Georgia. Email: temury.zaqarashvili@iliauni.edu.ge}
\altaffiltext{2}{Departament de Matem\`{a}tiques i Inform\`{a}tica.
\\ Universitat de les Illes Balears, \\ E-07122 Palma de Mallorca,
Spain. Email: marc.carbonell@uib.es} \altaffiltext{3}{Departament de
F\'{\i}sica, Universitat de les Illes Balears, E-07122 Palma de
Mallorca, Spain. Email: ramon.oliver@uib.es,
joseluis.ballester@uib.es}

\begin{abstract}
Apart from the 11-year solar cycle, another periodicity around
155-160 days was discovered during solar cycle 21 in high energy
solar flares, and its presence in sunspot areas and strong magnetic
flux has been also reported.  This periodicity has an elusive and
enigmatic character, since it usually appears only near the maxima
of solar cycles, and seems to be related with a periodic emergence
of strong magnetic flux at the solar surface.  Therefore, it is
probably connected with the tachocline, a thin layer located near
the base of the solar convection zone, where strong dynamo magnetic
field is stored. We study the dynamics of Rossby waves in the
tachocline in the presence of a toroidal magnetic field and
latitudinal differential rotation. Our analysis shows that the
magnetic Rossby waves are generally unstable and that the growth
rates are sensitive to the magnetic field strength and to the latitudinal
differential rotation parameters. Variation of the differential rotation and
the magnetic field strength throughout the solar cycle enhance the
growth rate of a particular harmonic in the upper part of the tachocline around the maximum of the solar cycle. This harmonic is symmetric
with respect to the equator and has a period of 155-160 days. A rapid increase of the wave
amplitude could give place to a magnetic flux emergence leading to observed periodicities in solar
activity indicators related with magnetic flux.

\end{abstract}


\section{Introduction}\label{intro}

During solar cycle 21, a short periodicity between 152--158 days was
discovered in $\gamma$ ray flares \citep {Rieger84}, X ray flares
\citep{Rieger84, Dennis85, Bai87, Kile91, Dimitropoulou08}, flares producing energetic
interplanetary electrons \citep{Droge90}, type II and IV radio bursts
\citep{Verma91}, and microwave flares \citep{Bai85, Kile91}.  However,
this periodicity was absent during solar cycle 22 \citep{Kile91,
Bai92a, ozguc94}.

The periodicity has also been detected in indicators of solar activity
(sunspot blocking function, sunspot areas, ``active'' sunspot groups,
group sunspot numbers) which suggest that it is associated
preferentially with photospheric regions of compact magnetic field
structures \citep{Lean89, Lean90, pap90, carball90, Bou92, carball92,
Verma92, oliver98, ballester99, Krivo02}.  Probably, the most
important, and enigmatic, feature of the periodicity is that it
appears during epochs of maximum activity and that it occurs in
episodes of 1 to 3 years.

\citet{rabin91} performed a study of the magnetic flux variations
during solar cycle 21 which reveals the existence of quasi-periodic
pulses or episodes of enhanced magnetic activity. The duration of
the pulses is $\approx$ 5 rotations during the years around maximum
activity, the epoch in which the flare periodicity appears, and the
comparison with magnetic field maps indicates that those pulses of
activity correspond to the occurrence of complex active regions
containing large sunspots \citep{Bai87a}.

\citet{ballester02, ballester04} analyzed several data sets of, or
strongly related to, photospheric magnetic flux to point out that the
appearance of the near 160-day periodicity in different manifestations
of solar activity during solar cycle 21 has its underlying cause in
the appearance of the periodicity in the magnetic flux linked to
regions of strong magnetic field.  They also showed that during solar
cycle 22 the periodicity does not appear in the photospheric magnetic
flux records and, as a consequence, the periodicity did not appear in
other solar activity indicators, while during solar cycle 23 it
appeared in the photospheric magnetic flux but not in other solar
activity indicators.

Several mechanisms have been put forward in order to explain the
existence of this periodicity.  \citet{wolff83} linked it to the
interaction of rotating features (active longitude bands) resulting
from g-modes with $l=2$ and $l=3$.  \citet{Bai87b} suggested that
the cause of this periodicity must be a mechanism that causes active
regions to be more flare productive.  Later, \citet{Bai87} concluded
that it cannot be due to the interaction of \lq \lq hot spots", i.e.
regions where flare activity is higher than elsewhere \citep{Bai87a,
Bai88}, rotating at different rates and that the cause must be a
mechanism involving the whole Sun.  \citet{Ichi85} suggested that it
is related to the timescale for storage and/or escape of magnetic
fields in the solar convection zone.  \citet{B90}, taking into
account the possible intermittency of the periodicity, suggested
that this behavior could be simulated with a damped, periodically
forced non-linear oscillator, which shows periodic behavior for some
values of the parameters and chaotic behavior for other values.
\citet{wolff92} argued that such periodicity can be understood in
terms of the normal modes of oscillation of a nearly spherical,
slowly rotating star, when two r-modes (inertial modes) couple with
an interior g-mode beat.  This suggestion seems to agree
qualitatively with the fact that the periodicity is stronger around
the activity maximum.  \citet{Bai91} and \citet{stu92} proposed that
the Sun contains a \lq \lq clock", modeled by an oblique rotator or
oscillator, with a period of 25.8 days and suggested that the
periodicity of 154 days is just a subharmonic of that fundamental
period.  Later, \citet{Bai93} modified the earlier period to the
value 25.50 days, but that model seems to be very constrained by
helioseismological data about the rotation of the Sun's interior.
\citet{lou00} suggested that such periodicities can be related to
large-scale equatorially trapped Rossby-type waves showing that, for
typical solar parameters, the periods of these waves (with n = 1 and
m even) are in good agreement with the observed ones. Moreover,
\citet{lou00} has also pointed out that such waves can give rise to
detectable features, such as surface elevations in the photosphere.
Coincidently, \citet{kuhn00} have reported observations made with
MDI onboard SOHO and claim to have detected a regular structure of
100-m \lq \lq hills'', uniformly spaced over the surface of the Sun
with a characteristic separation of 90,000 km.  They suggest that
this structure is the surface manifestation of Rossby waves, or
r-modes oscillations. Finally, \citet{Dimitropoulou08} have linked
the found periodicities in different classes (B, C, M, X) of X-ray
flares with the theoretical periods derived by \citet{lou00},
pointing out that odd m periodicities are also frequent and
significant.

On the other hand, most of the proposed mechanisms to explain solar
flares, specially the most energetic ones, accept as a prerequisite
the emergence of magnetic flux \citep{priest90, forbes91} which, by
reconnection with the ambient field,
triggers the destabilization of active regions.  Based on this
mechanism, Carbonell \& Ballester (1990, 1992) suggested that the
periodic increase in the occurrence rate of energetic flares is
related to a periodic emergence of magnetic flux through the
photosphere. Later, Oliver et al.  (1998) showed that during solar
cycle 21 there was a perfect time correlation between the intervals
of occurrence of the periodicity in sunspot areas and energetic
flares, and \citet{ballester02} clearly pointed out that in
cycle 21, and during the time interval in which the periodicity
appeared, there was a perfect time and frequency coincidence between
the impulses of high-energy flares and those corresponding to strong
photospheric magnetic flux. The efficiency of the reconnection
mechanism depends on the geometry of the two flux systems
\citep{Galsgaard07} and recent high resolution observations
performed by \citet{zuca08} have confirmed the suitability of the
mentioned mechanism for flare production.

Emerged magnetic flux is probably connected to deeper regions,
namely to the tachocline, which is a thin, transition layer between
differentially rotating convection zone and rigidly rotating
radiative envelope. The tachocline may prevent the spreading of the
solar angular momentum from the convection zone to the interior
\citep{spi92,gough98,gough07,gar07} and probably it is the place,
where the large-scale magnetic field which governs the solar
activity is generated/amplified.

The observed periodicity of 155--160 days in the emerging flux is in
the range of Rossby wave spectrum. Therefore, we suggest that the
periodicity is connected to the Rossby wave activity in the
tachocline. Rossby waves are well studied in the geophysical context
\citep{gill82,ped87}, however, the presence of magnetic fields
significantly modifies their dynamics \citep{zaqa07,zaqa09}. On the
other hand, the differential rotation, which is inevitably present
in the tachocline, may lead to the instability of particular
harmonics of magnetic Rossby waves. It has been shown that the joint
action of toroidal magnetic field and the differential rotation
generally leads to tachocline instabilities
\citep{gilman97,cal03,dik05,gil07,gil007}. However, the stability
analysis usually has been performed in an inertial frame, which
complicates to extract the information about unstable Rossby modes.
Therefore, it is of paramount importance to perform the stability
analysis in a rotating frame. Another important point is that the
consideration of a rotating frame may tighten the stability criteria
as it has been suggested by \citet{hug01}. The difference between the present
analysis and that by \citet{hug01} is the inclusion of
rotation which allows us to obtain Rossby wave solutions.

In this paper, we use a rotating spherical coordinate system to
study the linear stability of magnetic Rossby waves in the solar
tachocline taking into account the latitudinal differential rotation
and the toroidal magnetic field. We perform a two dimensional
analysis, which can be followed in the future by more sophisticated
shallow water considerations \citep{gil00}. We first derive the
analytical conditions of instability similar to \citet{dahlburg98}
and \citet{hug01}. Then, we perform  a detailed stability analysis
using Legendre polynomial expansions \citep{longuet68} to obtain the
spectrum of unstable harmonics of magnetic Rossby waves.


\section{Magnetic Rossby wave equations in the presence of differential rotation and the toroidal magnetic field}

Since the Rossby wave spectrum is clearly seen in the rotating
frame, in the following we use a spherical coordinate system $(r,
\theta, \phi)$  rotating with the solar equator, where $r$ is the
radial coordinate, $\theta$  is the co-latitude and $\phi$ is the
longitude.

The solar differential rotation law in general is
\begin{equation}\label{omega}
\Omega=\Omega_0 + \Omega_1(\theta),
\end{equation}
with
\begin{equation}\label{omega1}
\Omega_1(\theta)=-\Omega_0 (s_2 \cos^2 \theta + s_4 \cos^4 \theta),
\end{equation}
where $\Omega_0$ is the equatorial angular velocity, and $s_2, s_4$
are constant parameters determined by observations.

Rossby waves are mainly polarized in the plane perpendicular to
gravity, then a two-dimensional $(\theta, \phi)$ analysis is a good
approximation \citep{gill82}. The two-dimensional analysis is also
justified by Squire's theorem which states that for each unstable
3-dimensional disturbance there is a corresponding unstable
2-dimensional disturbance with stronger growth rate \citep{squ33}.

The magnetic field is predominantly toroidal, $\vec B = \Xi
\hat{e}_{\phi}$, in the solar tachocline, and we take $\Xi =
B_{\phi}(\theta) \sin \theta$, where $B_{\phi}$ is in general a
function of co-latitude. Then, the incompressible magnetohydrodynamic
(MHD) equations in the frame rotating with $\Omega_0$ are (see
appendix A):
\begin{equation} \label{MHD1}
{{\partial u_{\theta}}\over {\partial t}} +
\Omega_1({\theta}){{\partial u_{\theta}}\over {\partial \phi}} -
2[\Omega_0+\Omega_1({\theta})] \cos \theta u_{\phi} = -{{1}\over
{\rho R_0}}{{\partial p_t}\over {\partial \theta}}+ {{B_{\phi}}\over
{{4\pi\rho R_0}}}{{\partial b_{\theta}}\over {\partial \phi}}-2
{{B_{\phi} \cos \theta}\over {{4\pi\rho R_0 }}}b_{\phi},
\end{equation}
$$
{{\partial u_{\phi}}\over {\partial t}} +
\Omega_1({\theta}){{\partial u_{\phi}}\over {\partial \phi}}
+2\Omega_0 \cos \theta u_{\theta} + \Omega_1({\theta}) \cos \theta
u_{\theta} + u_{\theta}{{\partial }\over {\partial
\theta}}[\sin{\theta}\Omega_1({\theta})] =$$
\begin{equation}\label{MHD2}
 =-{{1}\over {R_0 \sin
\theta}}{{\partial p_t}\over {\partial \phi}}+{{B_{\phi}}\over
{{4\pi\rho R_0}}}{{\partial b_{\phi}}\over {\partial \phi}}+
{{b_{\theta}}\over {{4\pi\rho R_0 \sin \theta}}}{{\partial }\over
{\partial \theta}}(B_{\phi}\sin^2 \theta),
\end{equation}
\begin{equation}\label{MHD3}
{{\partial b_{\theta}}\over {\partial t}}+
\Omega_1({\theta}){{\partial b_{\theta}}\over {\partial \phi}} =
{{B_{\phi}}\over {{R_0}}}{{\partial u_{\theta}}\over {\partial
\phi}},\,\,\,\, {{\partial }\over {\partial \theta}}\left (\sin
\theta b_{\theta} \right ) + {{\partial b_{\phi}}\over {\partial
\phi}}=0,
\end{equation}
\begin{equation}\label{MHD4}
{{\partial }\over {\partial \theta}}\left (\sin \theta u_{\theta}
\right ) + {{\partial u_{\phi}}\over {\partial \phi}}=0,
\end{equation}
where $u_{\theta}$, $u_{\phi}$, $b_{\theta}$ and $b_{\phi}$ are the
velocity and magnetic field perturbations, $p_t$ is the  total
pressure (hydrodynamic plus magnetic), $\rho$ is the density and
$R_0$ is the distance from the solar center to the tachocline.

We consider the stream functions for velocity and magnetic field
\begin{equation}\label{stream}
u_{\theta}={1\over {{\sin \theta}}}{{\partial \Psi}\over {\partial
\phi}},\,\,u_{\phi}=-{{\partial \Psi}\over {\partial \theta}},\,\,
b_{\theta}={1\over {{\sin \theta}}}{{\partial \Phi}\over {\partial
\phi}},\,\,b_{\phi}=-{{\partial \Phi}\over {\partial \theta}}.
\end{equation}

Substitution of expressions (\ref{stream}) into
(\ref{MHD1})-(\ref{MHD4}) and Fourier analysis with $\exp[im(\phi -
c t)]$ gives
$$
(c-\Omega_1)\left [{{\partial }\over {\partial \theta}} \sin{\theta}
{{\partial }\over {\partial \theta}} - {m^2\over {\sin
\theta}}\right ]\Psi -2\Omega_0 \sin \theta \Psi + {d\over {d
\theta}}\left ( {1\over {\sin \theta}} {d\over {d \theta}}(\Omega_1
\sin^2 \theta)\right )\Psi=
$$
\begin{equation}\label{equation-momentum1}
= -{{B_{\phi}}\over {{4\pi\rho R_0}}}\left [{{\partial }\over
{\partial \theta}} \sin{\theta} {{\partial }\over {\partial \theta}}
- {m^2\over {\sin \theta}}\right ]\Phi + {1\over {4\pi\rho
R_0}}{d\over {d \theta}}\left ( {1\over {\sin \theta}} {d\over {d
\theta}}(B_{\phi} \sin^2 \theta)\right )\Phi,
\end{equation}
\begin{equation}\label{equation-induction1}
(c-\Omega_1)\Phi=-{{B_{\phi}}\over {{R_0}}}\Psi.
\end{equation}

Let us now make the transformation of variables $\mu=\cos \theta$, then
we obtain ($\Psi$ and $\Phi$ are normalized by $\Omega_0 R_0$ and
$B_0$ respectively, where $B_0$ is the value of $B_{\phi}$ at $\theta=0$)
\begin{equation}\label{momentum-last}
(\Omega_d - \omega)L\Psi + (2 - {d^2\over {d
\mu^2}}[\Omega_d(1-\mu^2)])\Psi - \beta^2 B L\Phi + \beta^2{d^2\over
{d \mu^2}}[B(1-\mu^2)] \Phi =0
\end{equation}
\begin{equation}\label{induction-last}
(\Omega_d-\omega)\Phi=B\Psi,
\end{equation}

where
$$
L={{\partial }\over {\partial \mu}}(1-\mu^2){{\partial }\over
{\partial \mu}} - {m^2\over {1-\mu^2}}
$$
is the Legendre operator and
$$\Omega_d(\mu)={\Omega_1(\mu)\over \Omega_0}, \,\, \omega={c\over \Omega_0},\,\,\, \beta^2={{B^2_0}\over
{{4\pi \rho \Omega^2_0R^2_0}}},\,\ B(\mu)={B_{\phi}(\mu)\over B_0}.
$$

Eqs. (\ref{momentum-last})-(\ref{induction-last}) govern the 2-dimensional
dynamics of magnetic Rossby waves in the presence of differential
rotation and toroidal magnetic field. The equations are analogous to
Eqs. (17)-(18) of \citet{gilman97}, but are written in the rotating
frame instead of in the inertial one.

\section{Analytical conditions of magnetic Rossby wave instability}

In this section, we derive the analytical instability bounds
using a well known technique
\citep{howard61,dra81,watson81,gilman97,dahlburg98,hug01}.

Let us define a new function $H$
$$
\Psi=(\Omega_d - \omega) H,\,\, \Phi= B H.
$$
Then Eqs. (\ref{momentum-last})-(\ref{induction-last}) can be cast
in the following form
\begin{equation}\label{equation-general}
{{\partial }\over {\partial \mu}}(1-\mu^2)P(\mu){{\partial H}\over
{\partial \mu}} - {m^2\over {1-\mu^2}}P(\mu)H +2(\Omega_d -
\omega)[1+(\mu \Omega_d)']H- 2\beta^2 B(\mu B)'H=0,
\end{equation}
where
$$
P(\mu)= (\Omega_d - \omega)^2 - \beta^2 B^2
$$
and $'$ means differentiation with respect to $\mu$.

Now, multiplying Eq. (\ref{equation-general}) by $H^*$, integrating
from -1 to 1 and using the boundary conditions $H(\mu=\pm 1)=0$, we
get

\begin{equation}\label{equation-inequality}
\int_{-1}^{1} {P(\mu)Q d \mu}-\int_{-1}^{1}{2(\Omega_d -
\omega)[1+(\mu \Omega_d)']|H|^2d \mu}+\int_{-1}^{1}{2\beta^2 B(\mu
B)'|H|^2d \mu}=0,
\end{equation}
where
$$
Q= (1-\mu^2)\left |{{\partial H}\over {\partial \mu}}\right |^2 +
{m^2\over {1-\mu^2}} |H|^2>0.
$$

Considering $\omega=\omega_r+i \omega_i$ in Eq.
(\ref{equation-inequality}) we obtain two different conditions for
instability (see detailed derivations in appendix B). The first
condition states that the instability takes place when

\begin{equation}\label{inequality1}
\omega_r^2 + \omega_i^2 \leq R^2_1,
\end{equation}
with
\begin{equation}\label{radius1}
R^2_1=\left [(s_2\mu^2+s_4\mu^4)^2 - \beta^2\mu^2 \right ]_{max}.
\end{equation}
In the remaining $_{max}$ and $_{min}$ mean maximal and minimal
values.

This means that the frequencies of unstable harmonics (actually
phase speeds, while frequencies can be obtained by multiplying by
$m$) lay inside the upper semicircle of complex $\omega$-plane with
center at the origin and radius $R_1$ (see Fig. 1).

The second instability condition is the semicircle theorem similar
to \citet{howard61}. The MHD generalization of Howard's semicircle
theorem in rectangular coordinates has been done by
\citet{dahlburg98} and \citet{hug01}. Here the theorem is derived in
the rotating spherical coordinate system as the second condition of
instability (see details in appendix B), obtaining

\begin{equation}\label{condition2}
\left (\omega_r - {{\Omega_{dmin}+\Omega_{dmax}}\over 2} \right )^2+
\omega_i^2 - \left ({{\Omega_{dmin}+\Omega_{dmax}}\over 2}\right )^2
+ \Omega_{dmin}\Omega_{dmax}- A_{max}\leq 0,
\end{equation}
where
\begin{equation}\label{A}
A(\mu)={{1-\mu^2}\over
{m^2}}(\Omega_{dmin}+\Omega_{dmax}-2\Omega_d)[1+(\mu \Omega_d)']
+{{1-\mu^2}\over {m^2}}2\beta^2 B(\mu B)' - \beta^2 B^2.
\end{equation}

We observe that $\Omega_{dmax}=0$ and $\Omega_{dmin}=-
\epsilon$, where $\epsilon=s_2 + s_4$, therefore we can write
\begin{equation}\label{semicircle}
\left (\omega_r + {{\epsilon}\over 2} \right )^2+ \omega_i^2 \leq
{{\epsilon^2}\over 4} + A_{max}.
\end{equation}

Due to this condition the frequencies of unstable modes lay inside
the semicircle of the complex $\omega$-plane with center
\begin{equation}\label{center}
\left ( -{{\epsilon}\over 2},0 \right )
\end{equation}

and radius (see Figure 1)

\begin{equation}\label{radius2}
R_2=\sqrt{{{\epsilon^2}\over 4} + A_{max}}.
\end{equation}

Equations (\ref{inequality1}) and (\ref{semicircle}) are two
necessary conditions of instability. They define two different
semicircles in the complex $\omega$-plane, and the instability occurs
when the two semicircles overlap (see \citet{hug01} for the same
statement in the rectangular case). If the radius of one semicircle
tends to zero, the instability disappears.

In the remaining we use a magnetic field
\begin{equation}\label{magnetic}
B_{\phi}=B_0 \mu,
\end{equation}
which changes sign at the equator \citep{gilman97}.

Now, we may estimate the instability bounds under tachocline
conditions. An important step is to choose the parameters of
differential rotation, $s_{2}$ and $s_{4}$. These parameters are
determined by observations and their values at the solar surface are
$s_{2} \approx s_{4} \approx 0.14$. Helioseismology shows that the
transition between the differentially rotating convective zone and
the rigidly rotating radiative interior is described by the function
$\Phi(r,r_c,w) = 0.5(1+erf[2(r-r_c)/w])$, where $erf$ is the error
function, $r_c$ is the radius of the central point of the tachocline
and $w$ is the characteristic thickness of the tachocline
corresponding to a variation of $\Phi(r)$ from $0.08$, at the bottom
of the tachocline, to $0.92$, at the tachocline's upper surface
\citep {koso96}. In order to calculate the parameters of the
differential rotation at the upper part of the tachocline, the solar
surface values must be multiplied by $0.92$, then, we obtain $s_2
\approx s_4 \approx 0.13$. However, it must be mentioned, that the real values of these parameters can be different in the tachocline \citep{charbonneau99} and also can change
through the solar cycle due to torsional oscillations \citep{labonte82,komm93,antia00,howe00,howe09}.
Therefore, these values are tentative and further observations are needed to infer the correct parameters and their cycle dependence.

The typical values of equatorial angular velocity, radius and
density in the tachocline are $\Omega_{0}= 2.7 \cdot 10^{-6}$
s$^{-1}$, $R_{0} = 5 \cdot 10^{10}$ \ cm and $\rho = 0.2$ \ g
$\cdot$ cm$^{-3}$ respectively. Then, the parameter $\beta^{2}$
is much smaller than unity being $\approx 0.0022$ for a magnetic
field strength of $10^{4}$ G. Using these parameters we get
$R_1=0.256$ and $R_2=0.154$ for azimuthal wave number $m=1$. Then,
the conditions (\ref{inequality1}) and (\ref{semicircle}) give that
the minimum period of the $m=1$ unstable modes in the tachocline is
\begin{equation}\label{minimum}
T_{min} \approx 105 \ {\rm days}.
\end{equation}
Therefore, only the magnetic Rossby modes with periods longer than
$105$ days may grow in time. However, equation (\ref{minimum}) only gives a lower bound
for oscillation periods. A more detailed analysis is required to
reveal the spectrum of possible unstable harmonics.

\section{Spectrum of unstable magnetic Rossby modes}

In this section, we use the general technique of Legendre polynomial
expansion \citep{longuet68}. Using the magnetic field profile
(\ref{magnetic}), Eqs. (\ref{momentum-last})-(\ref{induction-last})
are rewritten as

\begin{equation}\label{momentum-spectrum}
(\Omega_d - \omega)L\Psi + (2 - {d^2\over {d
\mu^2}}[\Omega_d(1-\mu^2)])\Psi - \mu \beta^2 L \Phi - 6\mu \beta^2
\Phi =0
\end{equation}
\begin{equation}\label{induction-spectrum}
(\Omega_d- \omega)\Phi=\mu \Psi.
\end{equation}

Let us expand $\Psi$ and $\Phi$ in infinite series of associated
Legendre polynomials
\begin{equation}\label{legandre}
\Psi=\sum^{\infty}_{n=m}a_nP^m_n(\mu),\,\,\,\Phi=\sum^{\infty}_{n=m}b_nP^m_n(\mu),
\end{equation}
which satisfy the boundary conditions $\Psi=\Phi=0$ at $\mu=\pm 1$.

The latitude-dependent part of the differential rotation has the form
\begin{equation}\label{omegad3}
\Omega_d=-s_2 \mu^2-s_4\mu^4.
\end{equation}

We substitute (\ref{legandre}) into Eqs.
(\ref{momentum-spectrum})-(\ref{induction-spectrum}) and, using a
recurrence relation of Legendre polynomials, we obtain algebraic
equations as infinite series (Details of the calculations can be
found in Appendix C for the case when the differential rotation has
only second order dependence on $\mu$ in expression
(\ref{omegad3})). The dispersion relation for the infinite number of
harmonics can be obtained when the infinite determinant of the
system is set to zero. In order to solve the determinant, we
truncate the series at $n=75$ and solve the resulting polynomial in
$\omega$ numerically.  The frequencies of different harmonics can be
real or complex giving the stable or unstable character of a
particular harmonic.  It turns out that $m=1$ harmonics are more
unstable such as it has been systematically shown by previous works
in many different occasions \citep{watson81,gilman97,dik05,gil07}.

Figure~\ref{fig2} shows the real, $mc_r$, and imaginary, $mc_i$,
frequencies of all $m=1$ unstable harmonics for different
combinations of differential rotation parameters and magnetic field
strength.  In order to show the dependence on the parameters
$s_2,s_4$, we vary these parameters for different values of magnetic
field strength so that the sum $s_2+s_4$ (which is the difference in
equatorial and polar angular velocities) remains 0.26.  In
Figure~\ref{fig2}, the upper left panel corresponds to the case
considered in Appendix C (i.e. $s_4=0$). Blue, green, yellow and red colors
correspond to magnetic field strengths of $2 \cdot 10^{3}$ \ G, $6
\cdot 10^{3}$ \ G, $2 \cdot 10^{4}$ \ G and $4 \cdot 10^{4}$ \ G,
respectively.  Asterisks (circles) denote the symmetric
(antisymmetric) harmonics with respect to the equator. The results
show that the $s_4 \mu^4$ term in the differential rotation
(\ref{omegad3}) significantly affects the behaviour of unstable
harmonics \citep{charbonneau99}. For each combination of $s_2,s_4$
and the magnetic field strength, there is a particular unstable
harmonic with a growth rate much stronger than for the other harmonics.
This harmonic is symmetric with respect to the equator and has the frequency of
0.17-0.18 $\Omega_0$ (yielding periods of 150-160 days) for the magnetic field strength of $\le 2 \cdot 10^{4}$ \ G.
The frequency decreases for stronger magnetic fields (red colors), therefore Rieger-type periodicities
arise as symmetric unstable harmonics for relatively weaker magnetic
field strength.

Thus, the appearance of a strong oscillation with a particular frequency needs a suitable
combination of differential rotation parameters ($s_2,s_4$) and
magnetic field strength. However, the differential rotation
parameters used in  Figure~\ref{fig2} are probably too high for the
solar tachocline. Therefore, we study the dependence of unstable
harmonics on more realistic differential rotation rates.

Figure~\ref{fig3} displays the dependence of the most unstable symmetric
harmonic (this harmonic can be identified on Fig.  2 as the blue,
green, yellow and red asterisks at top of each panel) on the differential
rotation parameters for two different values of the magnetic field.
Left panels correspond to the field strength of 2$\cdot 10^3$ \ G
and right panels correspond to the strength of $10^4$ \ G. Real and
imaginary parts of the harmonic vs $s_4$ are plotted for different
values of $s_2$.  The values of $s_2$ vary from 0.14 (blue color) to
0.09 (yellow color).  We can observe that the frequency, $m c_r$, of
this harmonic is only slightly dependent on the differential
rotation parameters and takes values between 0.16-0.18 $\Omega_0$
which correspond to oscillation periods of 150-170 days.  This is
the range where the Rieger-type periodicity has been observed.  On
the contrary, the growth rate, $m c_i$, of this harmonic strongly
depends on the differential rotation parameters.  The growth rate
becomes stronger when both $s_2$ and $s_4$, are increased.

The frequency and growth rate of this harmonic have no significant
dependence on the magnetic field when its strength is smaller than $
10^4$ \ G. Figure ~\ref{fig4} shows the dependence of the harmonic
calculated for three different profiles of the differential rotation
(blue line corresponds to $s_2=0.13$, $s_4=0.1$; the red line to
$s_2=0.11$, $s_4=0.12$ and green line to $s_2=0.11$, $s_4=0.1$).  We
can observe that the stronger growth rate occurs for the red line,
which means that $s_4$ is more important for the instability.

When the magnetic energy becomes comparable to the energy of differential rotation, then
the frequency of the symmetric harmonic is significantly reduced (see red asterisks on Figure ~\ref{fig2}).
The critical magnetic field strength, i.e. when the magnetic energy is comparable to the flow energy, is ${\sim} 5 \cdot 10^4$ \ G for the differential rotation parameters $s_2,s_4=$0.13. In this case, $(s_2 +s_4)^2 \sim \beta^2$, the radius of first semicircle $R_1$ (see Eq. (15)) tends to zero and the growth of symmetric unstable harmonics is suppressed.

\section{Discussion}

The periodicity of 155-160 days was discovered almost three decades
ago, however the reason of its appearance/disappearance is still
unknown. The most striking feature, perhaps, is its appearance only
at certain times, which normally coincide with the maximum of the
cycle (Figure~\ref{fig6}). This coincidence naturally suggests that
the magnetic field and the differential rotation at the solar cycle maximum
provide suitable conditions for the appearance of this periodicity.

Here we show that the periodicity can be connected to the dynamics
of magnetic Rossby waves in the tachocline, since, in this layer, they are
unstable due to the presence of toroidal magnetic field and
latitudinal differential rotation. First, we have derived the
analytical bounds of instability, which state that $m=1$ unstable
modes have periods $> 105$ days. Next, we have calculated the
detailed spectrum of unstable harmonics using the method of Legendre
polynomial expansion. We have found that the behaviour of
unstable harmonics is very sensitive to the combination of magnetic
field strength and the differential rotation parameters ($s_2,s_4$).
Each combination of the parameters favours a particular harmonic,
which has stronger growth rate compared to other unstable harmonics.
Therefore, this harmonic may quickly dominate over the others and may
lead to a detectable oscillation, if the parameters remain more or
less unchanged during some time. Unstable harmonics have two types
of symmetry with respect to the equator: symmetric and
antisymmetric. The growth rates
of symmetric modes are higher than the antisymmetric ones and they depend on the differential rotation parameters;
the growth becomes stronger for stronger shear.

Frequencies of symmetric unstable modes are in the range 0.16-0.18 $\Omega_0$ (Figure \ref{fig3}), which yield the periods of 150-170 days. In the case of strong differential rotation, their growth rate may reach up to 0.015 $\Omega_0$ i.e. the growth time is $\sim$ 280 days. Therefore, they may quickly dominate over the rest. The growth of the magnetic Rossby wave amplitude leads to an
enhanced magnetic buoyancy at the tachocline which causes the
periodic eruption of magnetic flux towards the solar surface.
Therefore, the periodicity is observed in the emerged magnetic flux
and consequently in many indicators of solar activity (see
references in the Introduction).

The question why the periodicities appear only at particular times (mostly just after solar maximum, see Figure~\ref{fig6})
needs additional explanation.  A possible reason is that the growth of symmetric harmonics strongly depends on the differential
rotation parameters ($s_2,s_4$).  It is known that the solar differential rotation is changing through the solar cycle.  The
pattern known as the torsional oscillation has been first observed at the solar surface in full disc velocity measurements
\citep{labonte82} and later in surface magnetic features as well \citep{komm93}.  Helioseismology shows that the torsional
oscillation is not only a surface phenomenon but may penetrate deeper into the solar interior
\citep{antia00,howe00,howe09}.  Then, the parameters $s_2,s_4$ may vary through the solar cycle in the tachocline, which
permits the strong growth of symmetric magnetic Rossby waves only at particular times.  This time should coincide with the
solar maximum.  We think that additional helioseismic estimations are needed to study this phenomenon.

One of the significant simplifications in our approach is the linear
stability analysis. The growth of perturbation amplitudes probably
leads to nonlinear effects. On the other hand, the process would be
accompanied by increased magnetic buoyancy, which causes the
eruption of magnetic flux upwards and consequently may stop further
growth of amplitudes. These processes should be studied with
sophisticated numerical simulations in the future.

It should be mentioned here that numerous previous papers have
studied the tachocline instabilities
\citep{gilman97,cal03,dik05,gil07,gil007}. However, all the
calculations have been performed in an inertial frame, while the
Rossby wave dynamics is more clearly seen in a rotating frame.
Another important difference between inertial and rotating frames is
that the instability conditions may be tightened in the moving frame
as suggested by \citet{hug01}.

The solar tachocline may consists of two parts: the inner radiative layer with a strongly stable stratification and the outer overshoot
layer with a weakly stable stratification \citep{gil00}.  The latitudinal differential rotation should be stronger in the
upper tachocline and weaker in the lower one.  On the contrary, the magnetic field strength should be higher in the lower
part and smaller in the upper one.  Therefore, the upper tachocline may favor the better conditions for the growth of
symmetric unstable harmonics, which trigger the Rieger-type periodicities.

\section{Conclusions}

In summary, we have shown that the destabilization of magnetic
Rossby waves in the solar tachocline is produced by the joint effect
of the latitudinal differential rotation and the toroidal magnetic
field. The frequencies and growth rates of unstable harmonics depend
on the combination of the differential rotation parameters and the magnetic
field strength.
The possible increase of latitudinal differential rotation at the solar maximum may trigger the instability of symmetric harmonic with period of 155-160 days in the upper part of the tachocline. This instability has a direct correlation with magnetic flux
emergence, therefore the periodicity also appears in solar
activity indicators related with magnetic flux. Later on, and
probably via reconnection, this periodic magnetic flux emergence
triggers the observed periodicity in solar flares.


The magnetic Rossby wave theory opens a new research area about the
activity on the Sun and other stars, and magnetic Rossby waves can
be of paramount importance for observed intermediate periodicities
in solar and stellar activity \citep{massi98, massi05}.

{\bf Acknowledgements} The authors acknowledge the financial support
provided by MICINN and FEDER funds under grant AYA2006-07637.  Also,
the Conselleria d'Economia, Hisenda i Innovaci\'o of the Government
of the Balearic Islands is gratefully acknowledged for the funding
provided under grant PCTIB2005GC3-03. T. V. Z. acknowledges
financial support from the Austrian Fond zur F\"orderung der wissenschaftlichen Forschung (under project P21197-N16), the Georgian National Science Foundation (under grant GNSF/ST06/4-098) and the Universitat de les Illes
Balears. Wavelet software was provided by C. Torrence and G. Compo
\footnote{The software is available at
http://paos.colorado.edu/research/wavelets}.

\appendix

\section{MHD equations in a rotating frame}

In the case of rigid rotation it is straightforward to transform
equations from inertial into the rotational frame, but the presence
of differential rotation slightly complicates the considerations as
different parts of the Sun rotate with different angular velocity.
The best way to overcome the difficulty is to consider the frame
rotating with the equator. Then the latitudinal differential
rotation can be considered as the unperturbed shearing motion in
this frame. 2-dimensional incompressible linearised MHD Equations ($\theta,
\phi$-plane) in the frame rotating with angular velocity of the
equator, $\Omega_0$, are
\begin{equation}\label{momentum-rotating1}
{{\partial u_{\theta}}\over {\partial t}} + {{U_{\phi}}\over {R_0
\sin \theta}}{{\partial u_{\theta}}\over {\partial \phi}} -
2\Omega_0 \cos \theta u_{\phi} - 2{{\cos \theta}\over {R_0 \sin
\theta}} U_{\phi}u_{\phi} = -{{1}\over {\rho R_0}}{{\partial
p_t}\over {\partial \theta}}+ {{\Xi}\over {{4\pi\rho R_0 \sin
\theta}}}{{\partial b_{\theta}}\over {\partial \phi}}-2 {{\Xi}\over
{{4\pi\rho R_0 }}}{{\cos \theta}\over {\sin \theta}}b_{\phi},
\end{equation}
$$
{{\partial u_{\phi}}\over {\partial t}} + {{U_{\phi}}\over {R_0 \sin
\theta}}{{\partial u_{\phi}}\over {\partial \phi}} +
{{u_{\theta}}\over R_0}{{\partial U_{\phi}}\over {\partial
\theta}}+2\Omega_0 \cos \theta u_{\theta} + {{\cos \theta}\over {R_0
\sin \theta}} u_{\theta} U_{\phi} =
$$
\begin{equation}\label{momentum-rotating2}
=-{{1}\over {R_0 \sin \theta}}{{\partial p_t}\over {\partial
\phi}}+{{\Xi}\over {{4\pi\rho R_0 \sin \theta}}}{{\partial
b_{\phi}}\over {\partial \phi}}+ {{b_{\theta}}\over {{4\pi\rho R_0
\sin \theta}}}{{\partial }\over {\partial \theta}}(\Xi \sin \theta),
\end{equation}
\begin{equation}\label{induction-rotating}
{{\partial b_{\theta}}\over {\partial t}}+ {{U_{\phi}}\over {R_0
\sin \theta}}{{\partial b_{\theta}}\over {\partial \phi}} =
{{\Xi}\over {{R_0 \sin \theta}}}{{\partial u_{\theta}}\over
{\partial \phi}},
\end{equation}
\begin{equation}\label{continuity-rotating}
{{\partial }\over {\partial \theta}}\left (\sin \theta u_{\theta}
\right ) + {{\partial U_{\phi}}\over {\partial \phi}}=0,
\end{equation}
\begin{equation}\label{maxwel-rotating}
{{\partial }\over {\partial \theta}}\left (\sin \theta b_{\theta}
\right ) + {{\partial b_{\phi}}\over {\partial \phi}}=0,
\end{equation}
where $u_{\theta}$, $u_{\phi}$, $b_{\theta}$ and $b_{\phi}$ are the
velocity and magnetic field perturbations, $\Xi$ and $U_{\phi}$ are
azimuthal components of unperturbed magnetic field and velocity in
the rotating frame, $p_t$ is the perturbation in total (hydrodynamic
plus magnetic) pressure.

We consider $U_{\phi}$ as the differential rotation with respect to
the equator, i.e.
\begin{equation}
U_{\phi}=R_0 \sin \theta \Omega_1({\theta}).
\end{equation}
Substitution of this expression into Eqs.
(\ref{momentum-rotating1})-(\ref{maxwel-rotating}) gives Eqs.
(\ref{MHD1})-(\ref{MHD4}).

\section{Derivation of analytical instability conditions}

The real and imaginary parts of Eq. (\ref{equation-inequality}) with
$\omega=\omega_r + i \omega_i$ are

\begin{equation}\label{equation-real}
\int_{-1}^{1}{\left [(\Omega_d - \omega_r)^2 - \omega_i^2 - \beta^2
B^2 \right ] Q d \mu}-\int_{-1}^{1}{2(\Omega_d - \omega_r)[1+(\mu
\Omega_d)']|H|^2d \mu}+\int_{-1}^{1}{2\beta^2 B(\mu B)'|H|^2d \mu}=0
\end{equation}
and
\begin{equation}\label{equation-imaginary}
2i \omega_i \left [ \int_{-1}^{1}{(\Omega_d - \omega_r)Q d \mu}
-\int_{-1}^{1}{[1+(\mu \Omega_d)']|H|^2d \mu} \right ]=0.
\end{equation}
Unstable harmonics should have non zero $\omega_i$, therefore Eq.
(\ref{equation-imaginary}) requires
$$
\int_{-1}^{1}{(\Omega_d - \omega_r)Q d \mu} =\int_{-1}^{1}{[1+(\mu
\Omega_d)']|H|^2d \mu}.
$$

Substitution of $\int_{-1}^{1}{\Omega_dQ d \mu}$ from this equation
into Eq. (\ref{equation-real}) leads to the equation

\begin{equation}\label{omega_d}
\int_{-1}^{1}{\left [\Omega_d^2 - \omega_r^2 - \omega_i^2 - \beta^2
B^2 \right ] Q d \mu}-\int_{-1}^{1}{2\Omega_d[1+(\mu
\Omega_d)']|H|^2d \mu}+\int_{-1}^{1}{2\beta^2 B(\mu B)'|H|^2d
\mu}=0,
\end{equation}
which then can be rewritten as
$$
\int_{-1}^{1}{\left [\Omega_d^2 - \omega_r^2 - \omega_i^2 - \beta^2
B^2 \right ] (1-\mu^2)\left |{{\partial H}\over {\partial
\mu}}\right |^2  d \mu} +
$$
$$
\int_{-1}^{1}{\left [\Omega_d^2 - \omega_r^2 - \omega_i^2 - \beta^2
B^2 - 2\Omega_d[1+(\mu \Omega_d)']{{1-\mu^2}\over m^2} + 2\beta^2
B(\mu B)'{{1-\mu^2}\over m^2}  \right ] {m^2\over {1-\mu^2}} |H|^2d
\mu} =0.
$$

This equation will be satisfied if both integrals are zero, which
requires

\begin{equation}\label{integral1}
(\Omega_d^2 - \beta^2 B^2)_{min} \leq \omega_r^2 + \omega_i^2
\leq(\Omega_d^2 - \beta^2 B^2)_{max}
\end{equation}
and
$$
\left (\Omega_d^2 - \beta^2 B^2 - 2\Omega_d[1+(\mu
\Omega_d)']{{1-\mu^2}\over m^2} + 2\beta^2 B(\mu B)'{{1-\mu^2}\over
m^2}\right )_{min} \leq \omega_r^2 + \omega_i^2 \leq
$$
\begin{equation}\label{integralB}
\left (\Omega_d^2 - \beta^2 B^2 - 2\Omega_d[1+(\mu
\Omega_d)']{{1-\mu^2}\over m^2} + 2\beta^2 B(\mu B)'{{1-\mu^2}\over
m^2}\right )_{max}.
\end{equation}

The inequality (\ref{integralB}) is similar to the inequality
(\ref{integral1}), but with two additional terms in the left and
right hand sides.  Both additional terms are positive,
therefore, inequality (\ref{integral1}) determines a condition of
instability. Using the profiles of magnetic field (\ref{magnetic})
and the differential rotation (\ref{omega1}), Eq. (\ref{integral1})
leads to Eq. (\ref{inequality1}) in the main text.

In order to obtain the semicircle theorem let us observe that

\begin{equation}\label{howard}
\int_{-1}^{1}{(\Omega_d - \Omega_{dmin})(\Omega_d - \Omega_{dmax})Q
d \mu} \leq 0.
\end{equation}

Then the substitution of $\int_{-1}^{1}{\Omega_d^2Q d \mu}$ from Eq.
(\ref{omega_d}) into Eq. (\ref{howard}) gives
$$
\int_{-1}^{1}{\left [\omega_r^2 + \omega_i^2 + \beta^2 B^2 -
(\Omega_{dmin}+\Omega_{dmax})\omega_r +
\Omega_{dmin}\Omega_{dmax}\right ] Q d \mu} \leq
$$
$$
\int_{-1}^{1}{(\Omega_{dmin}+\Omega_{dmax}-2\Omega_d)[1+(\mu
\Omega_d)']|H|^2d \mu}+\int_{-1}^{1}{2\beta^2 B(\mu B)'|H|^2d \mu}.
$$

This inequality can be rewritten as

$$
\int_{-1}^{1}{\left [\left (\omega_r -
{{\Omega_{dmin}+\Omega_{dmax}}\over 2} \right )^2+ \omega_i^2 +
\beta^2 B^2 - \left ({{\Omega_{dmin}+\Omega_{dmax}}\over 2}\right
)^2 + \Omega_{dmin}\Omega_{dmax}\right ](1-\mu^2)\left |{{\partial
H}\over {\partial \mu}}\right |^2d \mu}+
$$
$$
+\int_{-1}^{1}[\left (\omega_r - {{\Omega_{dmin}+\Omega_{dmax}}\over
2} \right )^2+ \omega_i^2 + \beta^2 B^2 - \left
({{\Omega_{dmin}+\Omega_{dmax}}\over 2}\right )^2 +
\Omega_{dmin}\Omega_{dmax}-
$$
\begin{equation}\label{inequality2m}
- {{1-\mu^2}\over
{m^2}}(\Omega_{dmin}+\Omega_{dmax}-2\Omega_d)[1+(\mu \Omega_d)']
-{{1-\mu^2}\over {m^2}}2\beta^2 B(\mu B)'] {m^2\over {1-\mu^2}}
|H|^2d \mu \leq 0.
\end{equation}

At least, one of the two integrals should have negative sign,
therefore
\begin{equation}\label{inequality20}
\left (\omega_r - {{\Omega_{dmin}+\Omega_{dmax}}\over 2} \right )^2+
\omega_i^2 + (\beta^2 B^2)_{min} - \left
({{\Omega_{dmin}+\Omega_{dmax}}\over 2}\right )^2 +
\Omega_{dmin}\Omega_{dmax}\leq 0
\end{equation}
and/or
\begin{equation}\label{inequality2B}
\left (\omega_r - {{\Omega_{dmin}+\Omega_{dmax}}\over 2} \right )^2+
\omega_i^2 - \left ({{\Omega_{dmin}+\Omega_{dmax}}\over 2}\right )^2
+ \Omega_{dmin}\Omega_{dmax}-A_{max}\leq 0,
\end{equation}
where
\begin{equation}\label{A}
A(\mu)={{1-\mu^2}\over
{m^2}}(\Omega_{dmin}+\Omega_{dmax}-2\Omega_d)[1+(\mu \Omega_d)']
+{{1-\mu^2}\over {m^2}}2\beta^2 B(\mu B)' - \beta^2 B^2.
\end{equation}

The inequality (\ref{inequality2B}) is wider than
(\ref{inequality20}). Therefore, it determines the second condition
of instability (Eq. (\ref{condition2}) in the main text).

\section{Derivation of dispersion equations using Legendre polynomial expansion}

The substitution of (\ref{legandre}) into Eqs.
(\ref{momentum-spectrum})-(\ref{induction-spectrum}) and using the
Legendre equation $LP^m_n + n(n+1)P^m_n=0$ leads to
$$
-(\Omega_d - \omega)\sum^{\infty}_{n=m}n(n+1)a_nP^m_n + (2 -
{d^2\over {d \mu^2}}[\Omega_d(1-\mu^2)])\sum^{\infty}_{n=m}a_nP^m_n
+ \mu \beta^2 \sum^{\infty}_{n=m}n(n+1)b_nP^m_n  -
$$
\begin{equation}
-6 \mu \beta^2\sum^{\infty}_{n=m}b_nP^m_n =0,
\end{equation}
\begin{equation}
(\Omega_d-\omega)\sum^{\infty}_{n=m}b_nP^m_n=\mu
\sum^{\infty}_{n=m}a_nP^m_n.
\end{equation}

Now we take the explicit form of $\Omega_d=-\epsilon \mu^2$, then
\begin{equation}\label{momentum-series}
\sum^{\infty}_{n=m}[\omega n(n+1) +2 + 2\epsilon]a_nP^m_n + \epsilon
\sum^{\infty}_{n=m}[n(n+1)-12]a_n\mu^2P^m_n +
\beta^2\sum^{\infty}_{n=m}[n(n+1)-6]b_n \mu P^m_n=0,
\end{equation}
\begin{equation}\label{induction-series}
\sum^{\infty}_{n=m}a_n \mu P^m_n + \sum^{\infty}_{n=m}\omega
b_nP^m_n+ \epsilon \sum^{\infty}_{n=m}b_n\mu^2P^m_n=0.
\end{equation}
We use the recurrence relations between Legendre polynomials,
namely:
$$
\mu^2P^m_n=A_nP^m_{n-2}+B_nP^m_{n}+C_nP^m_{n+2},
$$
$$
\mu P^m_n=D_nP^m_{n-1}+E_nP^m_{n+1},
$$
where
$$A_n={{(n+m)(n+m-1)}\over {(2n+1)(2n-1)}},\,\,\,B_n={{(n-m)(n+m)}\over {(2n+1)(2n-1)}}+{{(n-m+1)(n+m+1)}\over
{(2n+1)(2n+3)}},
$$
$$
C_n={{(n-m+1)(n-m+2)}\over {(2n+1)(2n+3)}},\,\,\, D_n={{n+m}\over
{2n+1}},\,\,\, E_n={{n-m+1}\over {2n+1}}.
$$

Substitution of these relations into Eqs.
(\ref{momentum-series})-(\ref{induction-series}) gives

$$
\sum^{\infty}_{n=m}[\omega n(n+1) +2 + 2\epsilon]a_nP^m_n +
\epsilon\sum^{\infty}_{n=m}[n(n+1)-12]A_na_nP^m_{n-2} +
\epsilon\sum^{\infty}_{n=m}[n(n+1)-12]B_na_nP^m_{n}+
$$
$$
+\epsilon\sum^{\infty}_{n=m}[n(n+1)-12]C_na_nP^m_{n+2}+
\beta^2\sum^{\infty}_{n=m}[n(n+1)-6]D_nb_nP^m_{n-1}+
\beta^2\sum^{\infty}_{n=m}[n(n+1)-6]E_nb_nP^m_{n+1}=0,
$$
$$
\sum^{\infty}_{n=m}a_nD_nP^m_{n-1}+\sum^{\infty}_{n=m}a_nE_nP^m_{n+1}
+ \sum^{\infty}_{n=m}\omega b_nP^m_n+ \epsilon
\sum^{\infty}_{n=m}A_nb_nP^m_{n-2}+ \epsilon
\sum^{\infty}_{n=m}B_nb_nP^m_{n}+ \epsilon
\sum^{\infty}_{n=m}C_nb_nP^m_{n+2}=0 .
$$

Rearranging terms we obtain

$$
\sum^{\infty}_{n=m}[\omega n(n+1) +2 + 2\epsilon]a_nP^m_n +
\epsilon\sum^{\infty}_{n=m}[(n+2)(n+3)-12]A_{n+2}a_{n+2}P^m_{n} +
\epsilon\sum^{\infty}_{n=m}[n(n+1)-12]B_na_nP^m_{n}+
$$
$$
+\epsilon\sum^{\infty}_{n=m}[(n-2)(n-1)-12]C_{n-2}a_{n-2}P^m_{n}+
\beta^2\sum^{\infty}_{n=m}[(n+1)(n+2)-6]b_{n+1}D_{n+1}P^m_n+
$$
$$
+\beta^2\sum^{\infty}_{n=m}[n(n-1)-6]b_{n-1}E_{n-1}P^m_n=0,
$$
$$
\sum^{\infty}_{n=m}a_{n+1}D_{n+1}P^m_n
+\sum^{\infty}_{n=m}a_{n-1}E_{n-1}P^m_n + \sum^{\infty}_{n=m}\omega
b_nP^m_n+ \epsilon \sum^{\infty}_{n=m}A_{n+2}b_{n+2}P^m_{n}+
\epsilon \sum^{\infty}_{n=m}B_nb_nP^m_{n}+
$$
$$
+ \epsilon \sum^{\infty}_{n=m}C_{n-2}b_{n-2}P^m_{n}=0 .
$$

Now the coefficients of $P^m_{n}$ give the equations
\begin{equation}\label{series1}
S_na_n + F_na_{n+2}+G_na_{n-2}+ H_nb_{n+1}+I_nb_{n-1}=0,
\end{equation}
\begin{equation}\label{series2}
J_na_{n+1} +K_na_{n-1}+ Q_nb_n +P_nb_{n+2} + M_nb_{n-2}=0,
\end{equation}
where
$$
S_n=\omega n(n+1) +2 + 2\epsilon +
\epsilon[n(n+1)-12]{{(n-m)(n+m)}\over
{(2n+1)(2n-1)}}+\epsilon[n(n+1)-12]{{(n-m+1)(n+m+1)}\over
{(2n+1)(2n+3)}},
$$
$$
F_n=\epsilon[(n+2)(n+3)-12]{{(n+m+2)(n+m+1)}\over
{(2n+5)(2n+3)}},\,\,\,
G_n=\epsilon[(n-2)(n-1)-12]{{(n-m-1)(n-m)}\over {(2n-3)(2n-1)}},
$$
$$
H_n=\beta^2[(n+1)(n+2)-6]{{n+m+1}\over {2n+3}},\,\,\,
I_n=\beta^2[n(n-1)-6]{{n-m}\over {2n-1}},\,\,\, J_n={{n+m+1}\over
{2n+3}},$$$$ K_n={{n-m}\over {2n-1}},\,\,\,Q_n=\omega +
\epsilon{{(n-m)(n+m)}\over
{(2n+1)(2n-1)}}+\epsilon{{(n-m+1)(n+m+1)}\over {(2n+1)(2n+3)}},
$$
$$
P_n=\epsilon{{(n+m+2)(n+m+1)}\over {(2n+5)(2n+3)}},\,\,\,
M_n=\epsilon{{(n-m-1)(n-m)}\over {(2n-3)(2n-1)}}.
$$

The expressions (\ref{series1})-(\ref{series2}) are infinite series and the dispersion relation for the infinite number of harmonics can be obtained when
the infinite determinant of the system is zero. In order to solve the determinant, we cut the series at $n=75$ and solve the resulting
polynomial in $\omega$ numerically.


\clearpage
\begin{figure}
\epsscale{0.6} \plotone{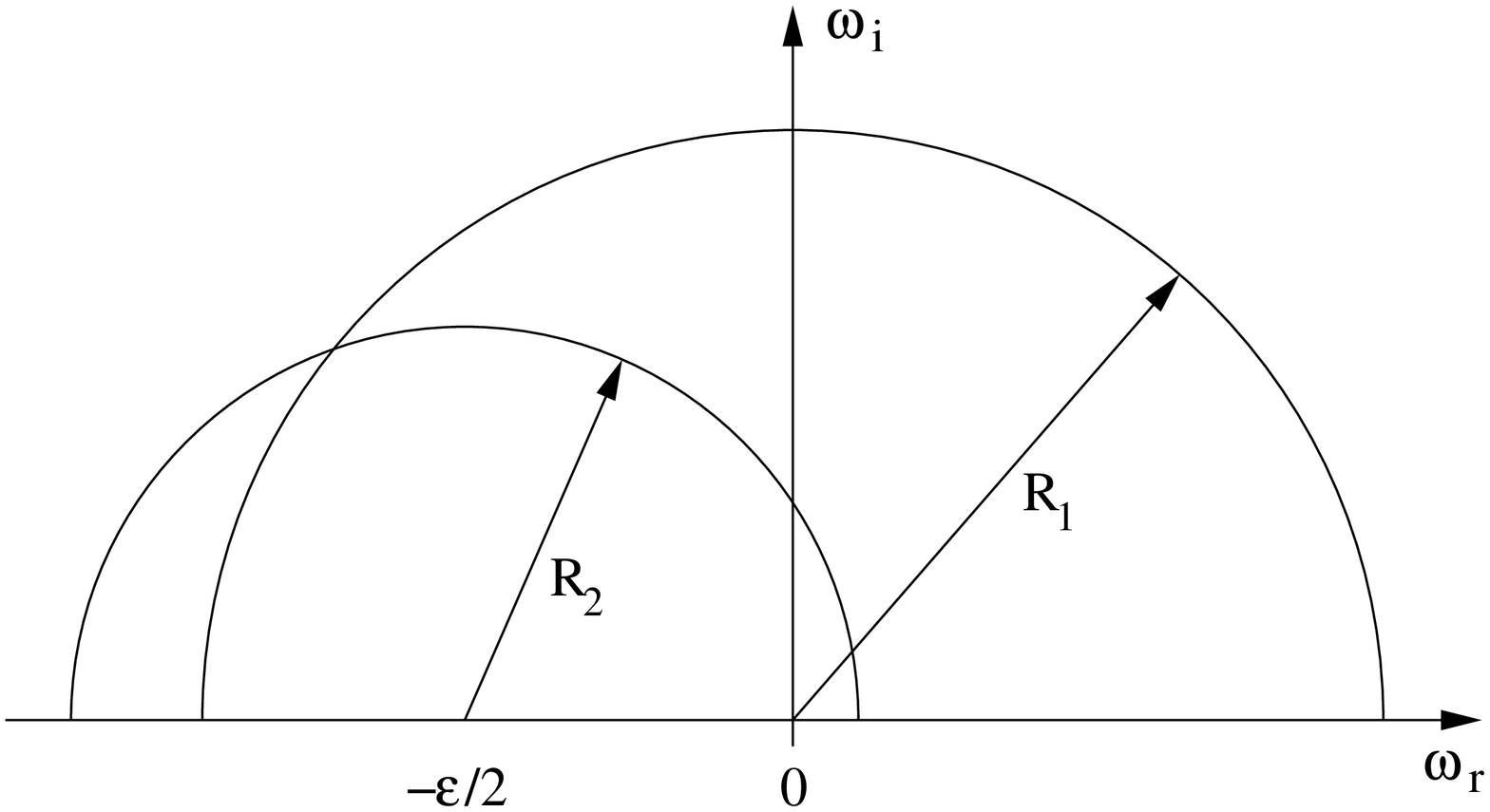} \caption{Semicircles of unstable
harmonics in the complex ($\omega_r$, $\omega_i$)-plane
corresponding to the two instability conditions, Eqs. (14) and (18).
Instability occurs when these two semicircles overlap. $\omega_r$,
$\omega_i$, $R_1$ and $R_2$ are normalised with respect to
$\Omega_0$.}\label{fig1}
\end{figure}

\clearpage

\begin{figure}
    \epsscale{1.1}
\plotone{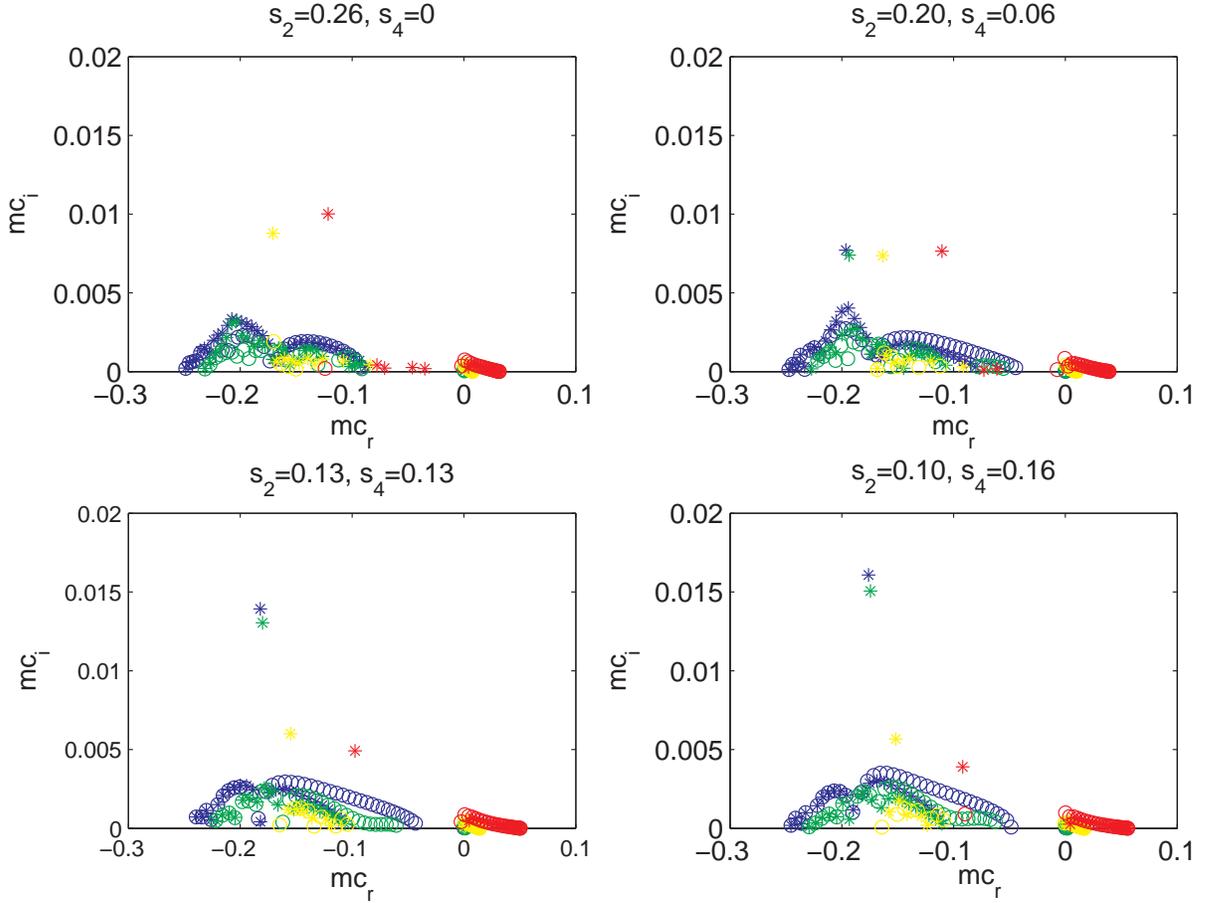}
 \caption{Real ($m c_r$) vs imaginary ($m c_i$) parts of unstable harmonic frequencies for different
 combinations of differential rotation parameters $s_2,s_4$ and magnetic field strengths (frequency is normalized by equatorial
 angular velocity, $\Omega_0$). Note that the difference between equatorial and polar angular velocities $s_2+s_4=0.26$ remains the same for all panels. The toroidal wave number $m$ equals 1.
 Blue, green, yellow and red colors
correspond to magnetic field strengths of $2 \cdot 10^{3}$ \ G, $6 \cdot 10^{3}$ \ G,  $2 \cdot 10^{4}$ \ G and $4 \cdot 10^{4}$ \ G, respectively.
Asterisks denote the symmetric harmonics with respect to the equator, while circles denote the antisymmetric ones. The frequencies are normalized by equatorial angular velocity, $\Omega_0$; for example, $m c_r=0.18$ corresponds to the period of $\sim$ 150 days.
} \label{fig2}
\end{figure}




\begin{figure}
\epsscale{1.1} \plotone{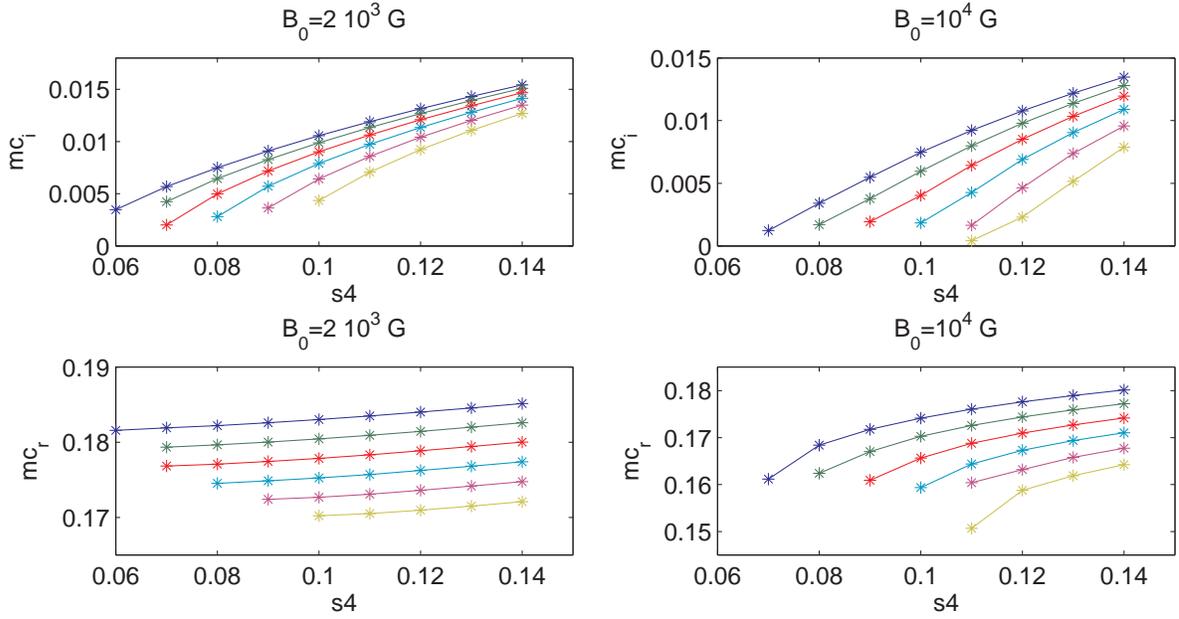} \caption{Real (lower panels)
and imaginary (upper panels) part of the frequency of the most unstable symmetric harmonic vs $s_4$ for
different values of $s_2$. Dark blue, green, red, blue, magenta and yellow colors correspond to   0.14, 0.13, 0.12, 0.11, 010 and 0.09  $s_2$  values respectively. The magnetic field strength equals to 2$\cdot 10^3$ \ G (left panels) and $10^4$ \ G
(right panel) respectively.} \label{fig3}
\end{figure}

\begin{figure}
\epsscale{0.8} \plotone{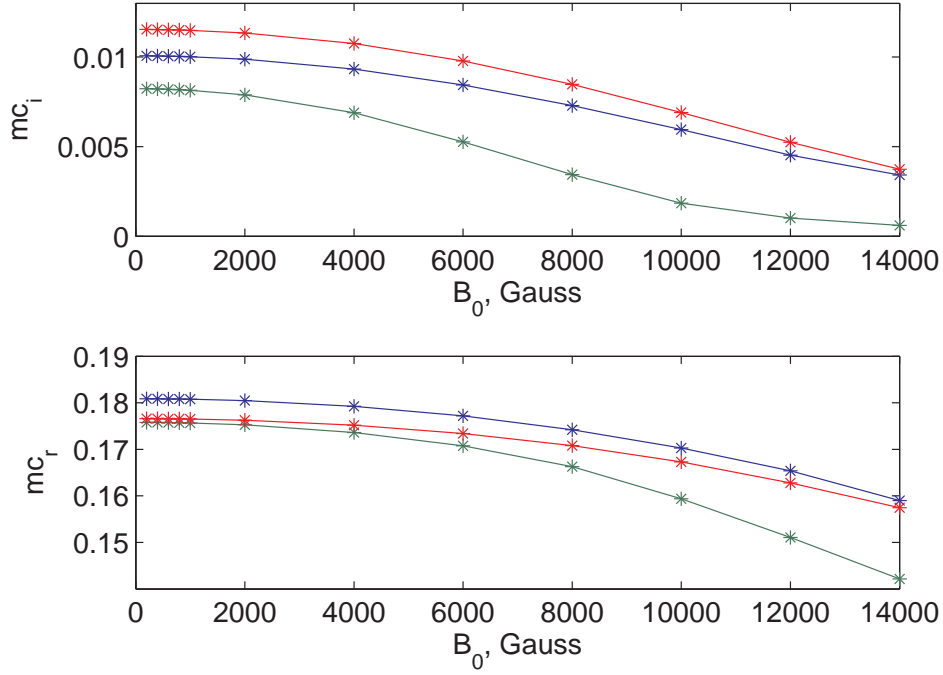} \caption{Dependence of real (lower panel) and
imaginary (upper panel) part of the frequency of the most unstable symmetric harmonic on the magnetic field
strength for three different combination of differential rotation parameters. The blue, green and red lines correspond to ($s_2$=0.13,$s_4$=0.1), ($s_2$=0.11, $s_4$=0.1) and ($s_2$=0.11, $s_4$=0.12) respectively.} \label{fig4}
\end{figure}

%

\begin{figure}
\epsscale{0.6} \plotone{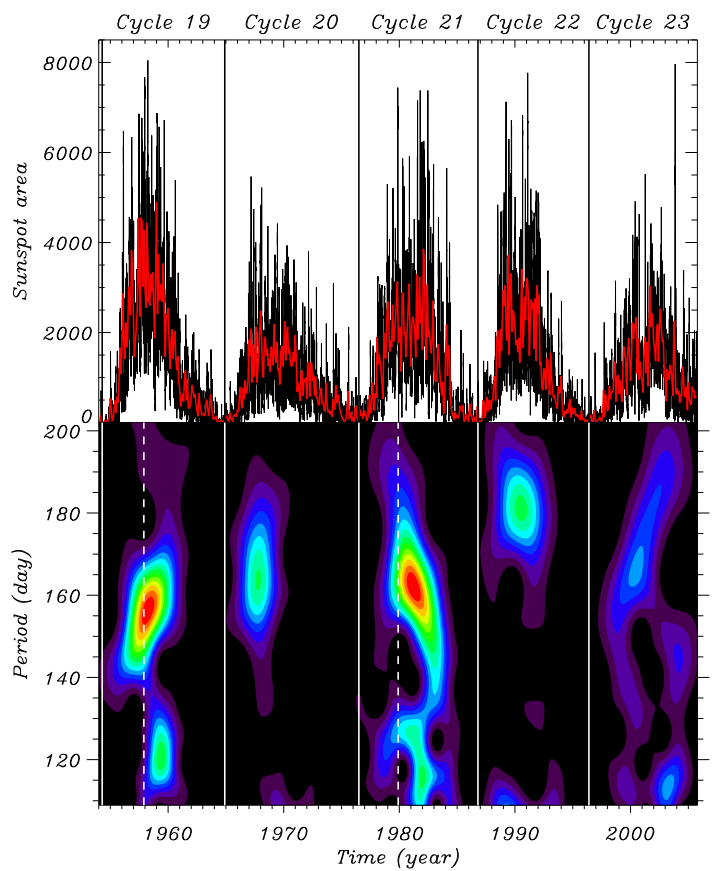} \caption{Top Panel: Plot of the
daily (black) and monthly averaged (red) sunspot areas for solar
cycles $19$ to $23$.  Bottom Panel: Time/period diagram calculated
using the Morlet wavelet \citep{To98} with $k_{0} = 20$. Vertical
solid white lines mark the epochs of minimum solar activity, while
the two dashed lines correspond to the maximum of cycles $19$ and
$21$. Large power values around 160 days can be seen in cycles 19,
20, 21 and 23, peaking at cycle 19.  Power is given in arbitrary
units.} \label{fig6}
\end{figure}

\end{document}